# Novel concept for pulse compression via structured spatial energy distribution

Venkata Ananth Tamma[1], Alexander Figotin[2] and Filippo Capolino[1]

1) Department of Electrical Engineering and Comp. Science, University of California Irvine, CA 92697
2) Department of Mathematics, University of California Irvine, CA 92697

**Abstract:** We present a novel concept for pulse compression scheme applicable at RF, microwave and possibly to optical frequencies based on structured energy distribution in cavities supporting degenerate band-edge (DBE) modes. For such modes a significant fraction of energy resides in a small fraction of the cavity length. Such energy concentration provides a basis for superior performance for applications in microwave pulse compression devices (MPC) when compared to conventional cavities. The novel design features: larger loaded quality factor of the cavity and stored energy compared to conventional designs, robustness to variations of cavity loading, energy feeding and extraction at the cavity center, substantial reduction of the cavity size by use of equivalent lumped circuits for low energy sections of the cavity, controlled pulse shaping via engineered extraction techniques. The presented concepts are general, in terms of equivalent transmission lines, and can be applied to a variety of realistic guiding structures.

## I. Introduction

Resonant cavities whose quality factor $Q$ can be modulated externally have been used to generate narrow pulses with very high peak power with applications in radars, linear accelerators and electronic counter-measure systems [1], [2]. Typically, microwave pulse compression (MPC) devices accumulate energy in a cavity with large $Q$ over an extended period of time $\tau_1$. After the accumulation of pre-determined amount of energy in the cavity, an external switching mechanism alters certain structural parameters of the cavity. This significantly reduces the $Q$ causing rapid release of the accumulated energy within a dump time $\tau_0$ [1], [2], [3] thus defining a pulse in time. The peak power of the outgoing pulse could be dramatically increased when compared to the power of the feeding source if $\tau_0 \ll \tau_1$ [1], [2], [3]. Previously, MPC devices based on the above operating principle, also known as active MPC devices [4], have been demonstrated [5]-[15]. Typically, a coupling mechanism such as an inductive iris has been used to couple energy into a resonant cavity and energy coupled out of such MPC devices by use of an externally activated switch [3], [5]-[15]. A common aspect among the previous implementations is the use of a resonant cavity, typically a conventional n($\lambda$/2) cavity resonator, formed by a transmission line (TL) or a waveguide.

Since MPC devices use cavities to store energy for long time durations, the quality factor $Q$ of the cavity plays an important role. The cavity $Q$ is limited by the loss of the stored energy via coupling to input/output and/or as heat due to skin-effect (Ohmic loss) in the metallic cavity walls [3]. Various techniques are known to impedance match a generator with a resonant cavity; see Chap. 6 in [16] and Chap. 4 in [17]. However, in general, it must be kept in mind that the unloaded $Q$ of the cavity is always reduced by the loading by the source impedance, hence limiting the capability to store energy. Hence, it is important to identify cavity structures with large unloaded $Q$ which does not drastically change upon loading by generator impedance.

Recently, it was shown that the stored energy in a conventional cavity with the length equal to a multiple of half-wavelength was essentially independent of the cavity length while the extractable instantaneous power was inversely proportional to the cavity length [8], [9]. This presents an important design trade-off between pulse amplitude and pulse length in resonant cavity based tunable MPC devices [8], [9]. The trade-off is due to the constant distribution of stored time-averaged energy inside a conventional n($\lambda$/2) cavity resonator shown schematically in Fig. 1 (a) in which we plot the spatial distribution of the sum of stored time average electric and magnetic energy density (units of J/m) in a conventional n($\lambda$/2) cavity resonator of length $L$ (numerical values are discussed in Sec. IV). Therefore, for a given amount of stored energy in the cavity, active MPC devices using such conventional cavities have a trade-off between the output pulse-width and output pulse power [8], [9]. One can overcome this trade-off constraint by designing devices in which energy is accumulated within a fraction of the cavity volume allowing for extraction of maximum amount of energy in the minimal possible time. Such a structured energy distribution, presented in this paper, is schematically shown in Fig. 1 (b) in which we plot the spatial distribution of the sum of stored time average electric and magnetic energy density (units of J/m) for a novel structured cavity having same length $L$ as the conventional n($\lambda$/2) cavity resonator. In contrast to the constant energy distribution in the conventional cavity shown in Fig. 1 (a), we find that most of the energy in Fig. 1 (b) is stored with a small region around the cavity center. This allows releasing the stored energy faster due to the reduction of the effective cavity length. Indeed, as detailed in Sec. II and IV, about 60 % of the total stored energy is concentrated within just 25 % of the cavity length.

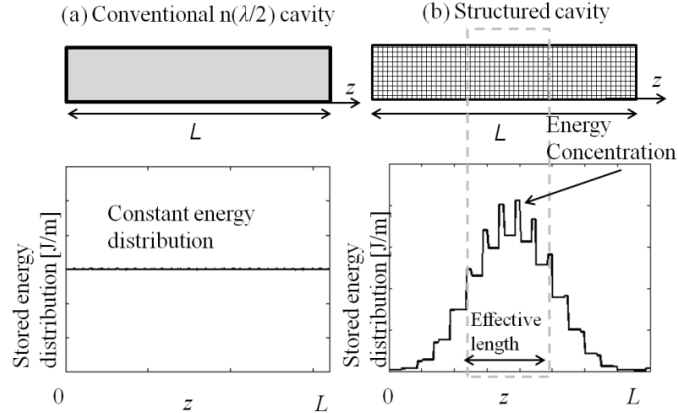

Fig. 1: Plots of spatial distribution of total time-average energy units of [J/m] versus position $z$ in (a) conventional n($\lambda$/2) cavity of length $L$ (b) structured cavity of length $L$. In (b), about 60 % of the total stored energy is stored in just 25 % of the total cavity length.

Such concentrated energy distribution is also helpful to control losses in structured cavity thereby improving the cavity $Q$. It is well known that the skin-effect loss is frequency dependent, see Chap. 1 in [16] and Chap. 2 in [17] and at lower frequencies, typically below 1 GHz, the skin-effect loss could be neglected, see Chap. 1 in [16] and Chap. 2 in [17]. It is possible to use cavities whose walls are coated with a thin layer low loss metal like gold or silver to reduce skin-effect losses. Since the energy in a conventional n($\lambda$/2) cavity resonator is uniformly distributed, to control losses the entire cavity has to be coated with gold or silver. However, a structured energy distribution allows to engineer the losses in only a small region of the cavity reducing them and improving performance.

We propose in this paper a novel structured cavity (i.e., with structured energy distribution) with applications to MPC devices. The cavity is composed of a cascade of $N$ unit cells, properly designed, each supporting two distinct modes (four, if distinguishing between forward and backward modes), thereby leading to distributed storage of energy in the cavity. The cavity $Q$ is very large and importantly it is *insensitive to loading* by source impedances. Such a cavity presents a novel feature compared to a standard cavity whose $Q$ is dramatically reduced when loaded by source impedance. A crucial aspect of the structured cavity is the distribution of a large percentage of energy within a small volume around the spatial center of the structure shown in Fig. 1 (b). This unique feature sets the structured cavity apart from standard resonant cavities which have a uniform distribution of energy. The properties of the new proposed structured cavities aid in efficient feeding and evacuation of accumulated energy. It also permits for substantial reduction in cavity size by allowing for lumped circuit implementation of those unit cells with lower stored energy. In addition, the cavity $Q$ could be further increased by reducing the losses only in those unit cells where most of the energy is stored. Another unique feature of the structured cavity is the preservation of spatial energy distribution around the cavity center with increasing number of unit cells $N$. Investigations into potential applications of many features of this novel cavity are already underway.

The novel features of the structured cavity are due to degenerate band-edge (DBE) modes [18]-[20]. At the DBE frequency, degenerate fourth power dependence of the radian frequency $\omega$ on the Bloch wavevector $k$ at the band-edge was demonstrated. This leads in particular to vanishing of the group velocity and consequent dramatic increase in the field intensity inside a finite stack exhibiting the DBE mode [18]-[20]. The concept of DBE [18]-[20] was first proposed in a simple structure composed of stacked anisotropic dielectric layers [18]-[20], where the importance of symmetry breaking in achieving the DBE was emphasized. Planar circuits supporting DBE modes were developed in analogy to propagation of light in the stacked anisotropic dielectric layers [21] [22]. Previously, it has been shown that field intensity enhancement varied as the fourth power of the number of unit cells *N* within the stack [18]-[20] and was found to be true only for large *N* [23]. The finite stack supporting DBE modes is ideally suited for highly frequency selective applications such as resonant cavities due to the narrow line shapes associated with the transmission band-edge resonance as evident from [18]-[20]. Such gigantic field enhancements and low group velocity are well suited for applications for the DBE modes in cavity structures for energy storage applications and high power generation.

Many unique features of the DBE mode supported by finite stacks studied in [18]-[20] can occur in structured resonant cavities. Indeed, as shown in [18]-[20], a very large *Q* was obtained by simply terminating the finite stack of anisotropic layers with vacuum. While various different cavity configurations are conceivable, in this paper we study a particular cavity configuration in which we feed and extract energy from the cavity center. As in [18]-[20], the basic configuration of the structure consists of a finite periodic stack of unit cells but terminated here in short circuits instead of vacuum as was done in [18]-[20]. In line with the definition of active MPC devices [1], [2], [3], [9] we show that the *Q* of this novel structured cavity can be dramatically altered by simple modifications to the structure thereby enhancing its appeal for use in MPC applications. This paper is organized as follows. In Sec. II, we describe the structured cavity made up of a finite number of unit cells *N* and discuss key attributes of the cavity such as quality factor and density of stored energy. In Sec III, we describe the unit cell design which is made up of periodic multiple transmission lines (MTLs). In this work, we model the structured cavity using cascaded sections of MTLs. It is understood that the equivalent TL model can be an exact field representation of complex realistic waveguiding systems [24], [25], [26]. In Sec. IV, we present an illustrative implementation of the unit cell of Sec. III and describe two states of the structure with dramatic differences in the quality factors. In addition, we discuss applications of the structured cavity to MPC devices. Although the concept of structured cavity is being introduced with MPC applications, we expect that it can be applied to printed or integrated RF circuits and optical devices.

**II. Structured Energy Distribution in a Cavity**

A schematic of the structured cavity of length $L = Nd$, consisting of *N* cascaded unit cells each of length *d* is shown in Fig. 2, in terms of equivalent transmission lines, where we denote the unit cell as $\mathbf{U}_n$, with, *n*=1,2, .., *N*. A unit cell can be formed by a few possible constituents and by several TLs, though in the rest of the paper we focus on having only two TLs. In all cases, there should be coupling between the two TLs within the unit cell and there should be a symmetry breaking within the cell in the propagation length and between the two TLs. Details of a specific unit cell are presented in Sec. III although other configurations are also possible. We note the lines at $z = \pm L/2$ are terminated in short circuits, though other load terminations would not alter the properties discussed here. We choose for convenience an even number of unit cells with the cavity centered and fed at $z = 0$ although we expect similar behavior for odd number of unit cells. We define $V_{S1}, V_{S2}$ and $Z_{S1}, Z_{S2}$ to be the source voltages and series source impedances feeding the upper and lower sets of TLs in the cavity and located at $z = 0$, though other source locations would be possible, even at the extremities of the cavity $z = \pm L/2$. In this paper, we assume a state-vector of the form $\mathbf{\psi}(z) = \begin{bmatrix} V_1(z) & V_2(z) & I_1(z) & I_2(z) \end{bmatrix}^T$ where, $V_1(z), I_1(z)$ and $V_2(z), I_2(z)$ are the voltages and currents at a point *z* on the upper and lower sets of TLs in the cavity. Throughout this paper, all TLs are assumed to have losses represented by series line resistances only which model the loss on the surface of metals in real waveguides. In particular, a constant TL distributed series line resistance $R_s = 1$ *m*Ω/m is used in all numerical calculations presented in this paper.

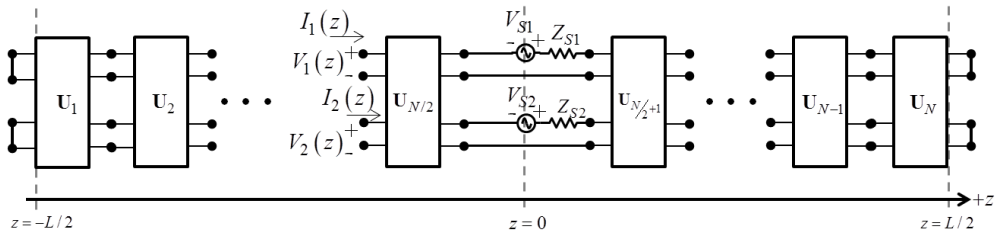

Fig. 2: Schematic of structured cavity formed by cascading *N* unit cells.

Similar to conventional cavities, the resonance frequency of the structured cavity can be computed by applying the transverse resonance method to the cavity at $z = 0$ and the method used is briefly discussed in Appendix B. For example, the resonant frequencies for structured cavities with $N$ = 8, 16 and 32 are $f_o$ = 4.684 GHz, 4.874 GHz and 4.883 GHz respectively and were calculated using the numerical parameters detailed in Appendix A. We first characterize the structured cavity by studying the spatial distributions of voltage, stored energy and energy loss per unit second along $z$ using the methods to compute them detailed in Appendices B and C. The results are obtained for the structured cavity fed by one ideal voltage source $V_{S1}$ =1 [V] and $V_{S2} = 0$. At any point $z$, we define $|V_{tot}(z)|^2 = |V_1(z)|^2 + |V_2(z)|^2$ as the total absolute squared voltage in the cavity and denote the maximum value of $|V_{tot}(z)|^2$ as $|V_{max}|^2$.

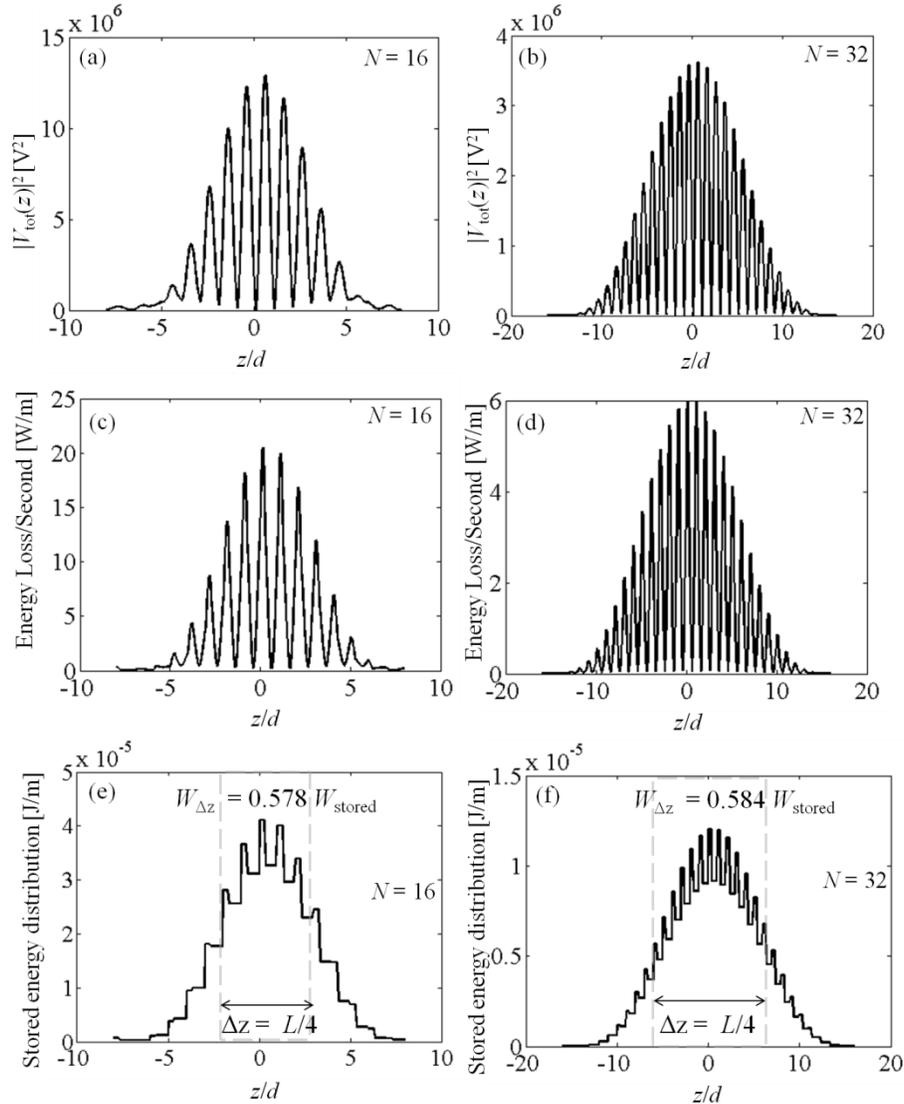

Fig. 3: Plots of $|V_{tot}(z)|^2$ versus position $z$ in a cavity with (a) $N$ = 16, and (b) $N$ = 32 unit cells. Plots of spatial distribution of energy loss per unit second versus position $z$ in a cavity in units of [W/m] with (c) $N$ = 16, and (d) $N$ = 32 unit cells. Plots of spatial distribution of time-average energy versus position $z$ in a cavity in units of [J/m] with (e) $N$ = 16, and (f) $N$ = 32 unit cells. In (c) – (f), about 58 % of the total stored energy is stored in just 25 % of the total cavity length.

In Fig. 3 (a, b), we plot $|V_{tot}(z)|^2$ as a function of $z, -L/2 \leq z \leq L/2$ in the structured cavity with $N = 16$ and $32$ unit cells respectively and realize that plots of $|V_{tot}(z)|^2$ are analogous to the field intensity ($|E|^2$ or squared amplitude) plots in [18]-[20]. For the sake of illustration we assume that each unit cell has nominal length $d=1$ [m]. In Fig. 3 (c, d) we plot the spatial distribution of the energy loss per unit second (power loss density) with units of [W/m] in the cavity with $N = 16$ and $32$ unit cells respectively using the formalism in Appendix C while in Fig. 3 (e, f), we plot the stored time-averaged energy (per unit length) with units of [J/m] in the cavity with $N = 16$ and $32$ unit cells respectively using the formalism in Appendix C. As expected, the spatial profiles of the energy loss per unit second and stored energy distribution in the cavity follow the same spatial profile as $|V_{tot}(z)|^2$. We note that the spatial profiles of squared voltage and total stored time-averaged energy are not located at the spatial center of the structured cavity ($z = 0$) and is attributed to the asymmetry in the unit cell and, in general, possible asymmetries in impedance terminations.

In both cavities with $N = 16$ and $32$ unit cells, $W_{stored}$ preserves the same spatial trend and the percentage of stored energy in sections of equal lengths of $L/4$ and centered on the geometric center of the cavity in both cavities is approximately equal. The cavity with $N = 16$ unit cells and total length of $L = 16d$ stores about 57.8 % of the total stored energy within a finite section of length $L/4$ centered around the geometric center of the cavity while the cavity with $N = 32$ unit cells and total length of $L = 32d$ stores about 58.4 % of the total stored energy within a finite section of length $L/4$ centered around the geometric center of the cavity.

The $Q$ of the structured cavity (called $Q_{cavity}$) is evaluated by use of the fundamental definition $Q = \omega_0 \text{(time-average energy stored)} / \text{(time-average energy loss / second)} = \omega_0 W_{stored} / P_{lost}$, where, $\omega_0 = 2\pi f_0$ is the radian frequency of resonance, see Chap. 6 in [16] and Chap. 7 in [17] and $P_{lost}$ is the total energy loss per unit second in the cavity. At their fundamental resonance frequencies, and assuming $Z_{S1} = Z_{S2} = 0$, the unloaded $Q$ of the cavity with $N = 16$ and $32$ unit cells are $Q_{cavity}(N=16) \approx 62000$ and $Q_{cavity}(N=32) \approx 62400$. Since the structured cavity is made up of TLs, we compare the $Q$ of the structured cavity with that of a standard short-circuited TL resonator formed by a TL of length $L = 16d$ corresponding to $31(\lambda/2)$ at a resonance frequency of 4.843 GHz which is very close to the resonant frequency of the structured cavity with $N = 16$ unit cells, $f_0 = 4.874$ GHz. Previously, standard short-circuited n($\lambda/2$)TL resonators have been explored in [3], [5]-[15] for MPC device applications. In [3], [5]-[15], the n($\lambda/2$) TL resonators are typically fed using an inductive iris on a metal wall located on the extremity of the resonator. In this work for simplicity the source impedance is considered purely resistive. We assume that the TL segment in the standard cavity has per unit length inductance and capacitance parameters as the maximum values of the distributed line inductances and capacitances of the MTLs whose numerical values are in Appendix A. As before, we account for losses by use of series line resistance and assume zero shunt conductance. At the resonance frequency, the unloaded $Q$ of the short-circuited n($\lambda/2$)TL resonator (called $Q_{TL}$) is calculated to be $Q_{TL} \approx 60000$ using well-known formulas Chap. 6 in [16] and Chap. 7 in [17]. Therefore, we observe then that for the same distributed series line resistance the unloaded $Q$ of the structured cavity is of the same order of magnitude as the unloaded $Q$ of the short-circuited n($\lambda/2$) TL resonator. It is important however to consider the effect of generator impedance loading on the $Q$ in both cases of the structured and standard cavities. Fig. 4 (a) shows the variation in $Q_{cavity}$ as a function of the purely resistive source impedance $Z_{S1}$ when the cavity is fed by $V_{S1}$ only and $Z_{S2} = 0$. Importantly, the quality factor $Q$ of the structured cavity is insensitive to different source impedances suggesting the robust nature of the cavity supporting DBE modes.

In Fig. 4 (b), we plot the loaded $Q$ of the standard n($\lambda/2$) TL resonator (called $Q_{TL}$) as a function of a purely resistive source impedance located at one of the extremities of the cavity. The loaded $Q_{TL}$ is seen to dramatically decrease, as expected, with increasing source impedance restricting the use of such resonators. Indeed, we find $Q_{TL} \approx 10$ when loaded by purely resistive source impedance $Z_{SRC} = 50$ [Ω]. The choice of feeding the TL resonator at one of the extremities was guided by the practical location of the inductive iris in the short-circuited n($\lambda/2$) TL resonator used in [3], [5]-[15]. However, similar trend in loaded $Q_{TL}$ is expected when the resonator is loaded by the source impedance at its center. In contrast, $Q_{\text{cavity}}$ is very stable when source impedance is varied with $Q_{\text{cavity}}(N=16) \approx 61500$ and $Q_{\text{cavity}}(N=32) \approx 62800$ when loaded by purely resistive source impedance $Z_{S1} = 50\,[\Omega]$, very close to their unloaded ($Z_{S1} = 0$) values and located at the center of the cavity. Therefore, considering loading by source impedance $Q_{\text{cavity}}$ is about two orders of magnitude larger than the loaded $Q$ of the standard n($\lambda$/2) TL resonator thereby enabling the structured cavity to store significantly more energy than conventional designs. Similar trend for $Q_{\text{cavity}}$ is observed when the cavity is loaded at one of its extreme ends.

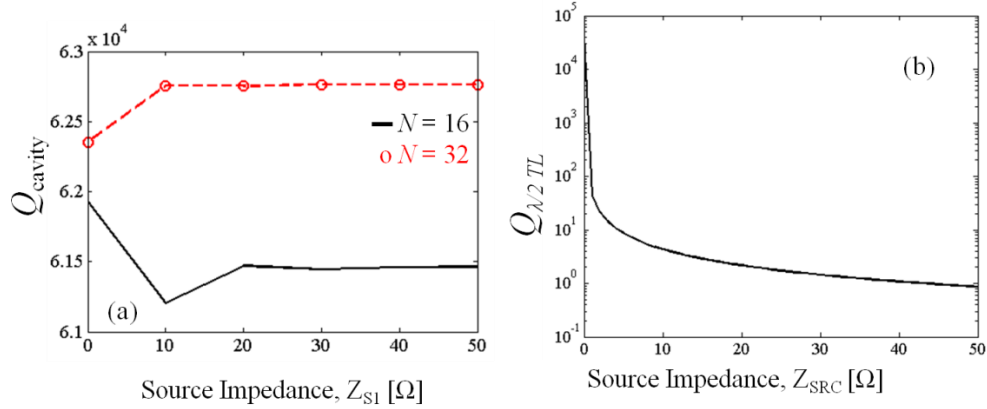

Fig. 4: Plots of variations in $Q$ of (a) a structured cavity of length $L = 16d$ and resonant at $f_0 = 4.8748\,\text{GHz}$, and (b) a standard $31(\lambda/2)$ TL resonator of same length $L = 16d$ and resonant at $f_0 = 4.8438\,\text{GHz}$, as a function of purely resistive source impedance.

In Fig. 3, we observe that the peak values of $|V_{\max}|^2$, the time-averaged stored energy and the energy loss per unit second in the structured cavity with $N = 32$ unit cells are all lower than those in the structured cavity with $N = 16$ unit cells. To further understand the behavior of the structured cavity, we plot the variation in $|V_{\max}|^2$ and $Q_{\text{cavity}}$ varying number of unit cells $N$, in Fig. 5.

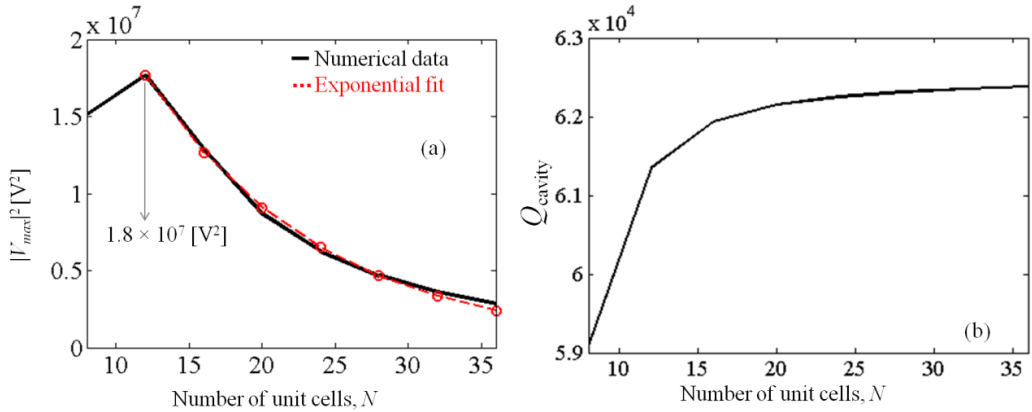

Fig. 5: Plots of (a) $|V_{\max}|^2$ along with an exponential fit and (b) $Q_{\text{cavity}}$ as function of number of unit cells, $N$.

In the structured cavity, the value of $|V_{\max}|^2$ is found to peak at $N = 12$ and thereafter exponentially reduce with increasing $N$ with the peak value at $N = 12$, $|V_{\max}|^2_{\text{peak}} = 1.8 \times 10^7\,[\text{V}^2]$. This decay behavior of $|V_{\max}|^2$ as a function of $N$ is in contrast with the previously reported results for field enhancement in finite stack [18]-[20] where the peak value of $|V_{max}|^2$ is asymptotically (for large $N$) proportional to $N^4$. Indeed, for values of $N \geq 12$, we observe that $|V_{max}|^2$ exponentially decays and can be fit by $1.8 \times 10^7 e^{-0.083(N-12)}$. Physically, the exponential reduction in $|V_{max}|^2$, though counter-intuitive (based on what said in [18]-[20]) can be attributed to the interplay between two competing processes: the enhancement in $|V_{max}|^2$ due to the DBE effect (discussed in [18]-[20]) and the energy lost due to losses in the TLs. One can see that losses dominate the enhancement due to the distributed nature of the cavity. The reduction in $|V_{max}|^2$ with increasing $N$ is also reflected in the values of $w_{stored}$ and $P_{lost}$ in the structured cavity. Indeed, the structured cavity with $N = 16$ unit cells is found to store about 28 % more energy than the cavity with $N = 32$ unit cells. However, $Q_{cavity}(N=32)$ is only marginally larger than $Q_{cavity}(N=16)$ despite the lower value of $|V_{max}|^2$ due to much lower $P_{lost}$ and on the whole, $Q_{cavity}(N)$ shows only a small increase with $N$. Notice that for large $N$ the energy $w_{stored}$ in the structured reduces with $|V_{max}|^2$ and therefore, we expect an optimal value of $N$ to obtain the best performance form the cavity. It should be noted that the large enhancement in $|V_{max}|^2$ experienced in this kind of cavities is not attributed to the use of short-circuit terminations since other loads would also prevent outflow of energy from the ends of the cavity. In other words, the resonance behavior here discussed would be preserved with a large variety of load terminations.

### III. Unit cell design and formalism

The design of the unit cell is fundamental to the operation of the structured cavity. The unit cells consists of two MTL segments A, B of lengths $d_A$ and $d_B$ shown schematically in Fig. 6 (a). Each MTL segment here consists of two TLs. These are chosen here to be uncoupled in segment A and coupled by distributed coupling capacitance in segment B, though many other configurations are possible and would lead to analogous results. The theoretical formulation for MTLs is well known and we follow the notation presented in [27]-[29]. We denote

$$\underline{\mathbf{Z}}_A = \underline{\mathbf{R}}_{s,A} + j\omega\underline{\mathbf{L}}_A \; , \; \underline{\mathbf{Y}}_A = j\omega\underline{\mathbf{C}}_A \qquad (1)$$

and

$$\underline{\mathbf{Z}}_B = \underline{\mathbf{R}}_{s,B} + j\omega\underline{\mathbf{L}}_B \; , \; \underline{\mathbf{Y}}_B = j\omega\underline{\mathbf{C}}_B \qquad (2)$$

to be the series impedance and shunt admittance matrices of segments A and B. Here, we define

$$\underline{\mathbf{R}}_{s,A} = \begin{bmatrix} R_{s,A1} & 0 \\ 0 & R_{s,A2} \end{bmatrix}, \; \underline{\mathbf{R}}_{s,B} = \begin{bmatrix} R_{s,B1} & 0 \\ 0 & R_{s,B2} \end{bmatrix} \qquad (3)$$

as the resistance matrices (with per unit length entries) for the segments A and B, respectively. We also define

$$\underline{\mathbf{L}}_A = \begin{bmatrix} L_{A,11} & 0 \\ 0 & L_{A,22} \end{bmatrix}, \; \underline{\mathbf{L}}_B = \begin{bmatrix} L_{B,11} & 0 \\ 0 & L_{B,22} \end{bmatrix} \qquad (4)$$

as the inductance matrices (with per unit length entries) for the segments A and B, respectively and

$$\underline{\mathbf{C}}_A = \begin{bmatrix} C_{A,11} & 0 \\ 0 & C_{A,22} \end{bmatrix}, \; \underline{\mathbf{C}}_B = \begin{bmatrix} C_{B,11}+C_{B,12} & -C_{B,12} \\ -C_{B,12} & C_{B,22}+C_{B,12} \end{bmatrix} \qquad (5)$$

as the capacitance matrices (with per unit length entries) for the segments A and B, respectively, where, $(L_{i,nn}, C_{i,nn})$ are the line inductance and capacitance, respectively, of the $n^{th}$ TL ($n = 1, 2$) in the $i^{th}$ ($i$ = A, B) segment, $C_{B,12} = C_{B,21}$ is the distributed coupling capacitance in segment B and $\omega$ is the radian frequency of operation. To achieve DBE regime, we note the requirement of symmetry breaking along both the horizontal and vertical axes in Fig. 6 (a). Here, we choose the two TLs in section A, $TL_{A1}$ and $TL_{A2}$, to be dissimilar and the two TLs in section B, $TL_{B1}$ and $TL_{B2}$, to be identical. Absence or presence of coupling breaks the symmetry between segments A and B. All transmission lines in sections A and B are assumed to have some losses represented by line resistances series which model the Ohmic loss on the surface of metals in real waveguides. We also assume lossless coupling between the two guided fields, represented by the two coupled TLs. The numerical values used in the computations are detailed in Appendix A.

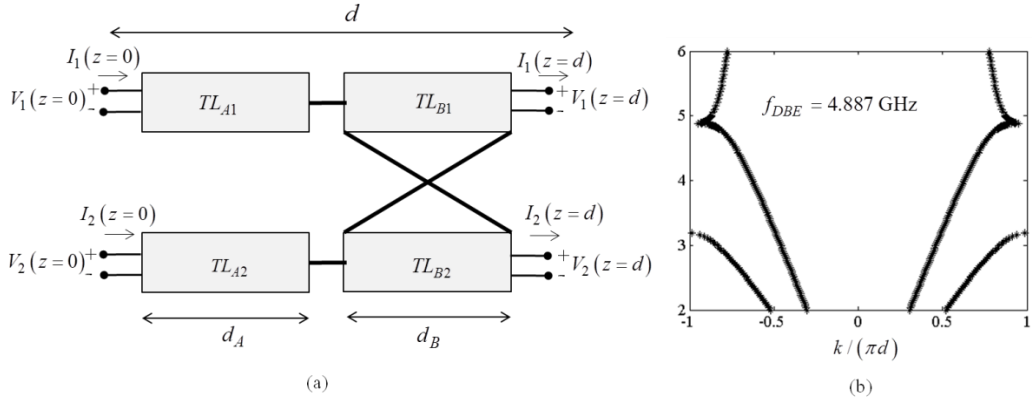

Fig. 6: (a) Schematic of a unit cell capable of supporting DBE modes. (b) Plot of the *k-ω* dispersion diagram for a periodic MTL cascading unit cells using the parameter values detailed in Appendix A.

Assuming the state-vector of the form $\underline{\psi}(z) = [V_1 \; V_2 \; I_1 \; I_2]^T$, the first order differential equations for the MTL in terms of the impedance and admittance matrix is written as

$$\frac{\partial}{\partial z}\underline{\psi}(z) = -\underline{M}\underline{\psi}(z), \qquad (6)$$

where $\underline{M} = \begin{bmatrix} \underline{0} & \underline{Z} \\ \underline{Y} & \underline{0} \end{bmatrix}$ and $\underline{Z}$ and $\underline{Y}$ are the impedance and admittance matrices describing the per unit parameters of the MTL. Denoting $\underline{\psi}(z_0) = \underline{\psi}_0$ as the initial TL values at $z_0$, assuming that $\underline{M}$ does not change with $z$ (i.e., for a uniform TL) we recognize (6) as the well-known Cauchy problem [18]-[20] with a unique solution $\underline{\psi}(z) = \underline{T}(z,z_0)\underline{\psi}(z_0)$, where, we define the matrix $\underline{T}(z,z_0)$, which uniquely relates the state vector $\underline{\psi}(z)$ between two known points $z_0$ and $z$, as $\underline{T}(z,z_0) = e^{-\underline{M}(z-z_0)}$.

Extending this concept to cascaded segments of MTL structures as in Fig. 6 (a), in the remainder of this paper it is convenient to resort to the definition of the "ABCD transfer matrix", see Chap. 4 in [16] and [17], commonly used in microwave engineering, and used here as generalized to multiple ports [27],[30]. We define the ABCD matrix for each section, A and B, of the 4 port circuits in Fig. 6(a) as $\underline{T}_A \equiv \underline{T}(z_0 - d_A, z_0) = e^{\underline{M}d_A}$, and $\underline{T}_B \equiv \underline{T}(z_0 - d_B, z_0) = e^{\underline{M}d_B}$ where,

$$\underline{M}_A = \begin{bmatrix} \underline{0} & \underline{Z}_A \\ \underline{Y}_A & \underline{0} \end{bmatrix}, \; \underline{M}_B = \begin{bmatrix} \underline{0} & \underline{Z}_B \\ \underline{Y}_B & \underline{0} \end{bmatrix} \qquad (7)$$

and $\underline{0}$ is a zero matrix of order 4. It is customary to implicitly assume $z < z_0$ in microwave engineering, see Chap. 4 in [16] and [17], and therefore the argument is dropped in the remaining of the paper. We can express the ABCD-like transfer matrix $\underline{\mathbf{T}}_U$ of the unit cell shown in Fig. 6 (a), as the product of two matrices describing the transfer matrices of the individual sections of the unit cell

$$\underline{\mathbf{T}}_U = \underline{\mathbf{T}}_A \underline{\mathbf{T}}_B. \tag{8}$$

For an infinitely long stack of TL unit cells, a periodic solution for the state vector $\mathbf{\psi}(z)$ exists in the Bloch form

$$\mathbf{\psi}(z+d) = e^{-jkd}\mathbf{\psi}(z), \tag{9}$$

where $k$ is the Bloch wavenumber. To determine the Bloch wavenumber, we can write the following eigenvalue equation

$$\underline{\mathbf{T}}_U \mathbf{\psi}(0) = e^{jkd}\mathbf{\psi}(0), \tag{10}$$

such that the four eigenvalues $\lambda_i = e^{jk_i d}$, $i = 1, 2, 3, 4$, of the $\underline{\mathbf{T}}_U$ operator are obtained as solutions of the characteristic equation

$$\mathrm{Det}\left(\underline{\mathbf{T}}_U - \lambda \underline{\mathbf{1}}\right) = 0, \tag{11}$$

where we define $\underline{\mathbf{1}}$ to be the identity matrix of order 4. The $k$-$\omega$ dispersion diagram, of the unit cell structure of Fig. 6 (a) is plotted in Fig. 6 (b), for the TL values given in Appendix A and is seen to exhibit the DBE mode at 4.887 GHz, at the edge of the Brillouin zone. The ABCD-like transfer matrix $\underline{\mathbf{T}}_U$ of the unit cell derived in this section is used in the formulation in Appendix B to calculate the ABCD-like transfer matrix of finite cascade of unit cells.

### IV. Illustrative implementation for storage and release of energy

We present here an illustrative implementation of the unit cell in Fig. 6 (a) using TLs and lumped capacitive coupling elements which readily demonstrates the possibility of $Q$ switching by breaking the DBE mode and thus strongly modifying the energy distribution and the cavity $Q$. The unit cell needs to be specifically designed so that the cavity supports the DBE mode with a large $Q$ value when in the 'On' state; whereas in the 'Off' state the DBE mode is destroyed by suitably designed structural modifications. We modify the unit cell in Fig. 6 (a) by replacing the distributed capacitance in segment B by a lossless lumped capacitor network as shown in Fig. 7 (a). Multiple switches are used within a unit cell as to switch the structure from 'On' state to 'Off' state. We assume ideal lossless switches with infinite off-state resistance, zero on-state resistance and zero switching time. The circuit of the modified unit cell in Fig. 7 (a) requires only lumped coupling between uncoupled TLs to create DBE mode and hence can easily be implemented using TEM-like waveguides, like coaxial cables, coupled by lumped capacitors or by real waveguides with stubs. The circuit in Fig. 7 (a) is in the 'On' state with the switch $S_M$ closed causing the TL segments A and B to be coupled, and with switches $S_{U1}$ and $S_{U2}$ open thereby decreasing the capacitive load of the circuit. The circuit in Fig. 7 (b) is in the 'Off' state with the switch $S_M$ open causing the TL sections A and B to be uncoupled from each other and with switches $S_{U1}$ and $S_{U2}$ closed thereby increasing the capacitive load of the two uncoupled and periodic upper and lower TL segments. These modifications provide for control over the wavenumber-frequency dispersion characteristic of the TL segments.

The transfer matrix $\underline{\mathbf{T}}_U$ of the unit cell in Fig. 7 (a), which uniquely relates the state vector $\mathbf{\psi}(z)$ between two known points $z$ and $z+d$, with $d > 0$, along the $+z$ axes such that $\mathbf{\psi}(z) = \underline{\mathbf{T}}_U \mathbf{\psi}(z+d)$, can be expressed as the product of three matrices describing the transfer matrices of the individual sections of the unit cell

$$\underline{\mathbf{T}}_U = \begin{bmatrix} \underline{\mathbf{C}}_A & \underline{\mathbf{S}}_A \\ \underline{\mathbf{S}}'_A & \underline{\mathbf{C}}_A \end{bmatrix} \begin{bmatrix} \underline{\mathbf{1}} & \underline{\mathbf{0}} \\ \underline{\mathbf{Y}}_l & \underline{\mathbf{1}} \end{bmatrix} \begin{bmatrix} \underline{\mathbf{C}}_B & \underline{\mathbf{S}}_B \\ \underline{\mathbf{S}}'_B & \underline{\mathbf{C}}_B \end{bmatrix}, \tag{12}$$

where we define **1** as the unit matrix of order 2, **0** as a zero matrix of order 2 and the admittance matrix
$\underline{\mathbf{Y}}_l = j\omega \begin{bmatrix} C_1 + C_m & -C_m \\ -C_m & C_2 + C_m \end{bmatrix}$. In addition, we define for the $i^{th}$ section, where $i=$ A, B, the following matrices

$$\underline{\mathbf{C}}_i = \begin{bmatrix} \cosh(\gamma_{i1}d_i) & 0 \\ 0 & \cosh(\gamma_{i2}d_i) \end{bmatrix}, \quad \underline{\mathbf{S}}_i = \begin{bmatrix} Z_i \sinh(\gamma_{i1}d_i) & 0 \\ 0 & Z_i \sinh(\gamma_{i2}d_i) \end{bmatrix}, \quad \underline{\mathbf{S}}'_i = \begin{bmatrix} \sinh(\gamma_{i1}d_i)/Z_i & 0 \\ 0 & \sinh(\gamma_{i2}d_i)/Z_i \end{bmatrix},$$

where, $\gamma_i$, $Z_i$ are the complex propagation constant and complex impedance for the $i^{th}$ section and whose values can be calculated from the RLGC physical parameters for the relevant segment [16],[17]. The transfer matrix $\underline{\mathbf{T}}_U$ of the unit cell derived in this section is used in the formulation in Appendix B to calculate the ABCD-like transfer matrix of finite cascade of unit cells.

The $k$-$\omega$ dispersion diagram of the unit cell in the 'On' state shown in Fig. 7 (a) is plotted in Fig. 7 (c) and is seen to exhibit the DBE mode at 3.310 GHz. However, in the 'Off' state, on uncoupling the circuit, we obtain two independent dispersion diagrams plotted in Fig. 7 (d) corresponding to the two uncoupled and periodic upper and lower TL segments. The difference between the $k$-$\omega$ dispersion diagrams corresponding to the 'On' and 'Off' states is clearly visible, showing in particular that the DBE dispersion phenomenon in Fig. 7 (c) has disappeared. We would like to stress that while we chose a simple uncoupling strategy, many other approaches can achieve the DBE disruption depending on the topology implementation of the structured resonant cavity.

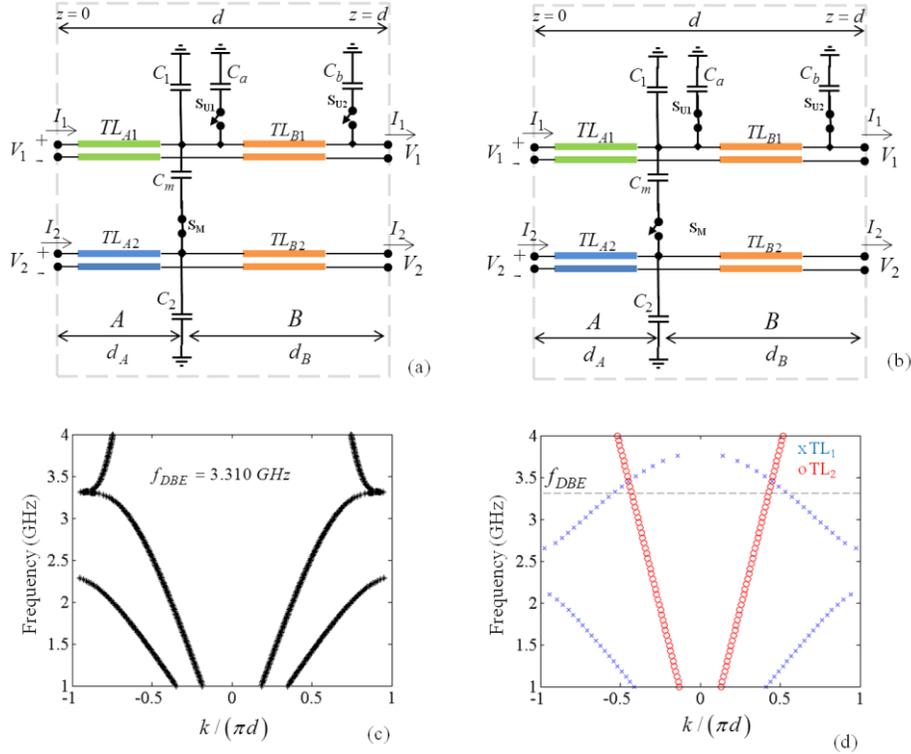

Fig. 7: Schematic of unit cell capable of supporting DBE mode in (a) 'On' state (b) 'Off' state. The $k$-$\omega$ dispersion diagram of the unit cell structure in (c) 'On' state and (d) 'Off' state, using the parameter values detailed in Appendix A.

As an illustrative example, a structured cavity was formed by cascading $N = 16$ unit cells in the 'On' state, unit cell shown in Fig. 7 (a), with short-circuit terminations at the four ports located at two extremities of the cavity. The cavity is fed at the spatial center by only one source $V_{S1} = 1$ [V] with $V_{S2} = 0$ and $Z_{S1} = Z_{S2} = 0$. We neglect $Z_{S1}$ since, as shown in Sec. II, it is not found to significantly affect the cavity performance. At the computed resonance frequency, $f_0 = 3.308$ GHz, we plot $|V_{tot}(z)|^2$ as a function of $z$ in Fig. 8 (a, b) using the formalism in Appendix B and also using a commercial circuit simulator AWR Microwave Office, respectively. The plots of $|V_{tot}(z)|^2$ obtained from circuit simulations are seen to be in excellent agreement with those obtained using the formalism in Appendix B. We state that the plots of $|V_{tot}(z)|^2$ in Fig. 8 (b) appears coarser than the corresponding plot in Fig. 8 (a) as only the voltages obtained at circuit nodes one per unit cell were plotted in Fig. 8 (b) whereas the voltages plotted in Fig. 8 (a) were obtained by sampling the entire cavity length with a discrete grid of spacing 0.05 m (20 sampling points per *d*).

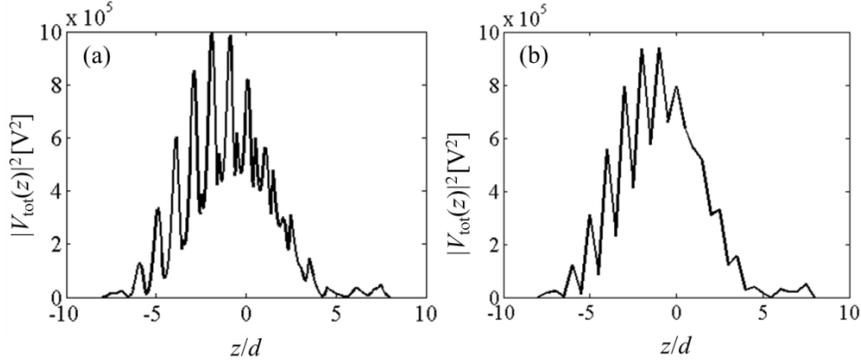

Fig. 8: Plot of $|V_{tot}(z)|^2$ versus position *z* for a cavity with *N* = 16 unit cells obtained using (a) the formalism in Appendix B, and (b) a circuit simulator.

The spatial energy distribution within the cavity with *N* = 16 unit cells in the 'On' state (unit cell shown in Fig. 7 (a)), is plotted in Fig. 9 at $f_0$ = 3.308 GHz using the formalism in Appendix C. In particular, in Fig. 9, the stored time-averaged energy in the TLs, which is a continuous curve in black color, is plotted separately from the stored time-average energy in the capacitive coupling network, which is discrete, in red color.

Of the total time-averaged energy stored in the cavity with *N* = 16 unit cells, about 99 % is stored in the TLs while the remaining 1 % is stored in the lumped capacitor network. From the schematic of the unit cell in 'Off' state, we observe that the energy stored in the lumped capacitor network cannot be extracted from the unit cell as the capacitor is disconnected from the network by the switch $S_M$ in open state. However, about 99 % of the total energy stored in the upper and lower cascaded segments of TLs and the loading capacitors can be extracted from the circuit by employing suitable extraction circuits.

Importantly, we find that an overwhelming fraction of the energy is stored in a small section of the cavity close to its center whereas sections of the cavity close to the extremities of the cavity store very little energy. Indeed, about 60 % of the total stored steady state energy is contained within just 25 % of the total length, or $\Delta z = 4d$ from $-3d \leq z \leq d$ and corresponding to four unit cells from $U_6$ to $U_9$, close to the spatial center of the stack. Such a distribution of energy is beneficial to both feed the cavity and extract energy from the cavity from its center and can lead to many advantages especially for applications in MPC devices.

Fig. 9 shows also a proposed energy extraction scheme in the form of a TL schematic which takes advantage of the spatial distribution of energy in the structured cavity. In particular, this scheme could make the structured cavity suited for MPC applications since it would be advantageous for MPC devices to produce large amplitude yet have very narrow pulse-width therefore delivering higher pulsed power. Since 60 % of the total energy is stored within four unit cells from $U_6$ to $U_9$, we choose to place switches such that the unit cells $U_6$ to $U_9$ are isolated from the rest of the circuit as shown in Fig. 9. These switches, which are normally closed, are open at the very same instant the unit cells are switched from 'On' to 'Off' state disrupting the DBE mode in the circuit and leading to isolated upper and lower TL segments as shown in Fig. 9. Indeed, it can be recognized that only units cells $U_6$ to $U_9$ need to be switched from 'On' to 'Off' state thereby reducing the number of switches required. In Fig. 9, we assume that the sources are automatically disconnected from the circuit using switches. In Fig. 9, we observe that energy can be extracted from the isolated upper and lower TL sections from multiple available open ports. Astute choices regarding the number of open ports from which energy can be extracted simultaneously can be made thereby generating pulses with very short pulse-widths. Since, the dispersion of the upper and lower TL segments can be effectively controlled by loading the circuit using capacitances $C_a$ and $C_b$, it could be possible to combine the energy simultaneously extracted from multiple ports with different phases thereby leading to efficient control over the pulse shape and width by use of engineered extraction techniques.

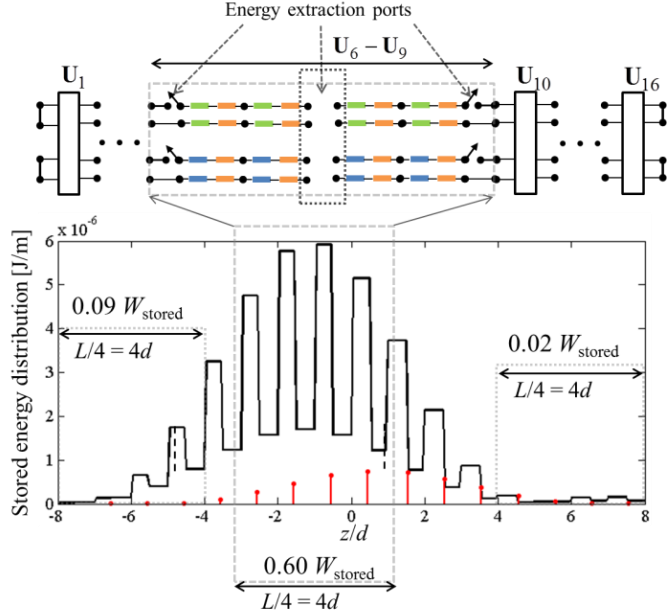

Fig. 9: Plot of the time-averaged stored energy distribution in the cavity with $N = 16$ unit cells showing that about 60 % of the total energy is stored in just 25 % of the total cavity length. Also, shown is a schematic of the energy extraction scheme with four unit cells in 'Off' state.

Since most of the energy is stored in units cells $U_6$ to $U_9$, $Q_{cavity}$ can be improved substantially by reducing losses in only these four unit cells. Indeed, for the structured cavity in Fig. 9 in 'On' state, reducing the series line resistance $R_s$ from 1 $m\Omega$/m to 0.1 $m\Omega$/m in just the 4 units cells $U_6$ to $U_9$ improves $Q_{cavity}$ by almost 100 % from 35000 to 68000 showing significant performance improvement can be achieved by further engineering of the structured cavity.

An important advantage of the structured energy distribution which is the basis for the proposed energy extraction scheme in Fig. 9, is the extraction of energy from the structured cavity with much narrower pulse-widths than conventional n($\lambda$/2) in resonant cavities. For example, we consider a n($\lambda$/2) resonant TL cavity having the same length as a structured cavity with $N = 16$ unit cells, $L = 16d$. For the case in which both the structured cavity and n($\lambda$/2) cavity have a port at the cavity center for energy feeding and extraction, the output pulse-width $\tau_0$ is proportional to $\Delta / v_g$, where, $v_g$ is the group velocity and $\Delta$ is the length of the cavity from which energy is extracted. Let $v_{g,U}$ be the group velocity in the upper TL segments in Fig. 9 and $v_{g,TL}$ be the group velocity in the n($\lambda$/2) cavity. Since the dispersion of the upper TL segments can be controlled by capacitive loading, for simplicity, we assume energy extraction from only the upper TL array and $v_{g,U} = v_{g,TL}$. Assuming the $Q$ of both the n($\lambda$/2) cavity and the structured cavity is switched after accumulation of same amount of time-averaged stored energy in both cavities and denoting $\tau_{0,TL}$ and $\tau_{0,U}$ to be the output pulse-widths from then ($\lambda$/2) and structured cavity, respectively, we obtain $\tau_{0,U} = \tau_{0,TL} / 4$ or the output pulse-width from a structured cavity is four times smaller than the pulse-width from a n($\lambda$/2) cavity.

Another advantage of the structured energy distribution is the possibility to substantially reduce the cavity size by use of equivalent lumped circuits in sections of the cavity with lower stored energy. We observe from Fig. 9 the net energy stored in regions of cavity from $-8d \leq z \leq -4d$ and $4d \leq z \leq 8d$ is only about 10 % of $W_{\text{stored}}$. Therefore, it could be beneficial to implement the TLs of unit cells in these regions by lumped iterative structures. For example, the TLs of unit cells $U_6$ to $U_9$ could be real waveguide segments while TLs in unit cells $U_1$ to $U_4$ and $U_{13}$ to $U_{16}$ could be implemented as lumped iterative structures consisting of cascade of many lumped inductors and capacitors since they only need to host a fraction of the whole structured cavity energy. Such a scheme could result in a physically smaller and lighter cavity with capability to store more energy when compared to conventional designs and can have many implications for MPC devices since the physical size and weight of cavities restrict mobile applications of MPC devices.

Finally, we studied the structured cavity for many different values of series distributed loss resistance $R_s$ and as expected, found $Q_{\text{cavity}}$ to be inversely proportional to it. We find that the structured cavity preserves all the observed spatial properties for several values of $R_s$. Due to the periodic arrangement of unit cells integral to its design, the structured cavity is scalable to any frequency with the losses only dependent on the implementation in that frequency. We anticipate that an all dielectric implementation could significantly reduce losses and improve the total stored energy and $Q_{\text{cavity}}$. Such implementations could potentially be useful also at optical frequencies and we expect that the concept of structured cavity could be applied to printed or integrated RF circuits and optical devices.

**V. Conclusions**

We proposed a novel cavity exhibiting structured energy distribution by properly cascading *N* unit cells, each satisfying symmetry breaking in two directions. We found the cavity *Q* to be very large and insensitive to loading by source impedances. This novel result is in contrast to the *Q* of a standard TL cavity which is dramatically reduced when loaded by source impedance. We showed that the structured cavity *Q* could be further increased by reducing the losses only in those unit cells where most of the energy is stored. It was also found that a large percentage of energy is distributed within a small region of the cavity around the spatial center of the cavity. These features allow for efficient feeding and faster evacuation of accumulated energy. Several key aspects of the cavity relevant to MPC applications were discussed. An illustrative implementation of the unit cell using TLs and lumped capacitor coupling network was presented. A proposal to switch the *Q* of the structured cavity was discussed and a possible energy extraction scheme allowing for narrow pulse-width generation was considered. In addition, efficient pulse shaping and pulse-width control can be achieved by engineering extraction schemes. The size of the structured cavity can be substantially reduced by lumped circuit implementation of those unit cells with lower stored energy. We anticipate many applications for the structured cavity in MPC devices, integrated RF circuits and optoelectronic devices.

**Appendix A: Parameter values used for numerical computations and circuit simulations**

In this paper, a constant distributed loss series line resistance $R_s = 1$ [$m\Omega$/m] is used in all numerical calculations. The distributed shunt conductance is always set to zero. All results are obtained for the structured cavity fed by one ideal voltage source $V_{S1} = 1$ [V] and $V_{S2} = 0$. All unit cells in this work have a nominal length $d$=1 [m]. The following are the parameters of the MTLs for the unit cell in Fig. 6 (a): (segments A and B are mentioned in subscripts) $d_A = 0.265$ [m], $d_B = 0.735$ [m], $L_{A,11} = L_{A,22} = 2$ [nH/m], $L_{B,11} = L_{B,22} = 2$ [nH/m], $C_{A,11} = 20$ [pF/m], $C_{A,22} = 2$ [pF/m], $C_{B,11} = C_{B,22} = 2$ [pF/m] and $C_{B,12} = C_{B,21} = 1.5$ [pF/m]. The following are the parameters of the TLs for the unit cell in Fig. 7 (a): $d_A = 0.495$ [m], $d_B = 0.505$ [m], $L_{A,11} = L_{A,22} = 2$ [nH/m], $L_{B,11} = L_{B,22} = 2$ [nH/m], $C_{A,11} = 20$ [pF/m], $C_{A,22} = 2$ [pF/m], and $C_{B,11} = C_{B,22} = 2$ [pF/m]. The following are parameters of the lumped capacitors: $C_1 = C_2 = 0.1$ [pF], $C_a = C_b = 5$ [pF] and $C_m = 3$ [pF]. We use 2 [nH/m] and 20 [pF/m] as the distributed line inductance and capacitance, respectively, of the standard short-circuited n($\lambda$/2) TL resonator.

**Appendix B: Transverse resonance method for cavity**

The procedure to obtain the resonance frequencies of the resonant cavity supporting DBE modes is briefly outlined. In this work, for simplicity, we consider feeding the cavity using only one ideal voltage source at a time, either of $V_{S1}$ or $V_{S2}$. In this section, we use the ABCD-like transfer matrix $\underline{\mathbf{T}}_U$ of the unit cells in Figs. 6 (a) and 7 (a) derived in Sec. III and IV respectively, to calculate the ABCD-like transfer matrix of finite cascade of unit cells. We begin with a general cavity structure with $N$ unit cells terminated in impedances $\underline{Z}_{Load,L1}$, $\underline{Z}_{Load,L2}$ on the left and $\underline{Z}_{Load,R1}$, $\underline{Z}_{Load,R2}$ on the right side, respectively as shown in Fig. A1 (a). For simplicity, we assume $N$ to be even and the cavity fed at $z = 0$. We combine the $N/2$ transfer matrices to the left and right of $z = 0$ corresponding to the chain of $N/2$ unit cells to the left and right of $z = 0$ and define the matrices $\underline{\mathbf{T}}_{Chain,R} = \prod_{n=N/2+1}^{N} \underline{\mathbf{T}}_U$ and $\underline{\mathbf{T}}_{Chain,L} = \prod_{n=N/2}^{1} \underline{\mathbf{T}}_U^{-1}$, that provide $\psi(-L/2) = \underline{\mathbf{T}}_{chain,L}\psi(0)$ and $\psi(0) = \underline{\mathbf{T}}_{Chain,R}\psi(L/2)$. Such a simplified representation of the cavity in Fig. A1 (a) is given by Fig. A1 (b). Compared to Fig. 2, we note that in Fig. A1 (a), the terminations are general load impedances rather than short circuits which is a special case treated in Sec. II.

After decomposing $\underline{\mathbf{T}}_{Chain,R}$ into $2 \times 2$ matrices such that

$$\underline{\mathbf{T}}_{Chain,R} = \begin{bmatrix} \underline{\mathbf{A}}_R & \underline{\mathbf{B}}_R \\ \underline{\mathbf{C}}_R & \underline{\mathbf{D}}_R \end{bmatrix}, \tag{13}$$

we write

$$\mathbf{V}_R(z=0) = \underline{\mathbf{A}}_R \mathbf{V}_R(L/2) + \underline{\mathbf{B}}_R \mathbf{I}_R(L/2) \tag{14}$$

$$\mathbf{I}_R(z=0) = \underline{\mathbf{C}}_R \mathbf{V}_R(L/2) + \underline{\mathbf{D}}_R \mathbf{I}_R(L/2) \tag{15}$$

where, $\mathbf{V}_R(z) = [V_{1,R}(z) \ V_{2,R}(z)]^T$, $\mathbf{V}_L(z) = [V_{1,L}(z) \ V_{2,L}(z)]^T$, $\mathbf{I}_R(z) = [I_{1,R}(z) \ I_{2,R}(z)]^T$ and $\mathbf{I}_L(z) = [I_{1,L}(z) \ I_{2,L}(z)]^T$.

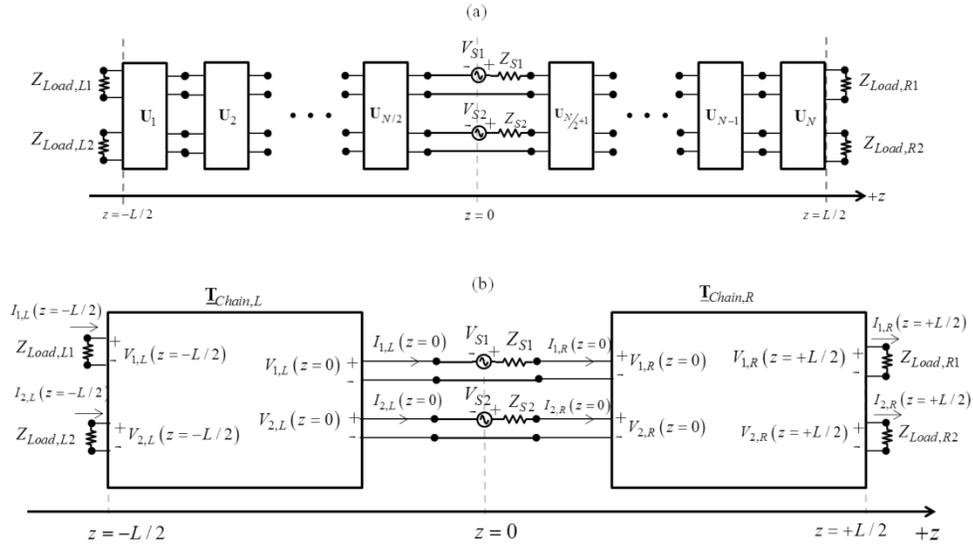

Fig.A1: (a) Generalized schematic of cavity with load impedances. (b) Simplified representation of Fig. A1 (a).

At $z = +L/2$, we note that $\mathbf{V}_R = \underline{\mathbf{Z}}_{Load,R} \mathbf{I}_R$, where,

$$\underline{\mathbf{Z}}_{Load,R} = \begin{bmatrix} Z_{Load,R1} & 0 \\ 0 & Z_{Load,R2} \end{bmatrix} \quad (16)$$

and hence (14), (15) are simplified to a two port representation at $z = 0$,

$$\mathbf{V}_R(0) = \underline{\mathbf{Z}}_{Chain,R} \mathbf{I}_R(0), \quad (17)$$

where, $\underline{\mathbf{Z}}_{Chain,R} = (\underline{\mathbf{A}}_R \underline{\mathbf{Z}}_{Load,R} + \underline{\mathbf{B}}_R)(\underline{\mathbf{C}}_R \underline{\mathbf{Z}}_{Load,R} + \underline{\mathbf{D}}_R)^{-1}$ represents the impedance finite length of the impedance terminated chain observed from the input terminals which are connected to the sources. Similarly, at $z = 0$ we obtain

$$\mathbf{V}_L(0) = \underline{\mathbf{Z}}_{Chain,L} \mathbf{I}_L(0), \quad (18)$$

where $\underline{\mathbf{Z}}_{Chain,L} = (\underline{\mathbf{A}}_L \underline{\mathbf{Z}}_{Load,L} + \underline{\mathbf{B}}_L)(\underline{\mathbf{C}}_L \underline{\mathbf{Z}}_{Load,L} + \underline{\mathbf{D}}_L)^{-1}$ and we decompose $\underline{\mathbf{T}}_{Chain,L}$ into $2 \times 2$ matrices such that

$$\underline{\mathbf{T}}_{Chain,L} = \begin{bmatrix} \underline{\mathbf{A}}_L & \underline{\mathbf{B}}_L \\ \underline{\mathbf{C}}_L & \underline{\mathbf{D}}_L \end{bmatrix}, \underline{\mathbf{Z}}_{Load,L} = \begin{bmatrix} Z_{Load,L1} & 0 \\ 0 & Z_{Load,L2} \end{bmatrix} \quad (19)$$

Denoting

$$\mathbf{V}_S = \begin{bmatrix} V_{S1} & V_{S2} \end{bmatrix}^T, \underline{\mathbf{Z}}_S = \begin{bmatrix} Z_{S1} & 0 \\ 0 & Z_{S2} \end{bmatrix}, \quad (20)$$

we apply Kirchhoff's Voltage law around both the source loops at $z = 0$ to obtain

$$\mathbf{I}_R(0) = \mathbf{I}_L(0) = (\underline{\mathbf{Z}}_{Chain,R} - \underline{\mathbf{Z}}_{Chain,L} - \underline{\mathbf{Z}}_S)^{-1} \mathbf{V}_S, \quad (21)$$

where, we assume that $(\underline{\mathbf{Z}}_{Chain,R} - \underline{\mathbf{Z}}_{Chain,L} - \underline{\mathbf{Z}}_S)$ can be inverted. Equations (14), (15) and (17), (18), (21) allow for evaluation of the state vectors $\mathbf{\psi}(z = 0^-) = \begin{bmatrix} V_{1,L}(z=0) & V_{2,L}(z=0) & I_{1,L}(z=0) & I_{2,L}(z=0) \end{bmatrix}^T$ and $\mathbf{\psi}(z = 0^+) = \begin{bmatrix} V_{1,R}(z=0) & V_{2,R}(z=0) & I_{1,R}(z=0) & I_{2,R}(z=0) \end{bmatrix}^T$. With these state vectors, combined with the definition of transfer matrices and the use of boundary conditions, the state vector $\mathbf{\psi}(z)$ at an arbitrary point $z, -L/2 \le z \le L/2$ in the cavity can be computed.

We compute the input impedance seen by the source for a particular case. If we assume $V_{S2} = 0$ and $Z_{S2} = 0$, the input impedance seen by the voltage generator $V_{S1}$, with $Z_{S1} = 0$, is $Z_{in,1} = V_{S1} / I_{1,L}(0)$ or $Z_{in,1} = V_{S1} / I_{1,R}(0)$. Using (17), (18) and (21), the input impedance seen by the voltage generator $V_{S1}$ is $Z_{in,1} = \det(\underline{\mathbf{Z}}_{diff}) / Z_{diff,22}$, where, $\underline{\mathbf{Z}}_{diff} = \underline{\mathbf{Z}}_{Chain,R} - \underline{\mathbf{Z}}_{Chain,L}$ and $Z_{diff,22}$ is the (2, 2) entry of the matrix $\underline{\mathbf{Z}}_{diff}$. Similarly, if we assume $V_{S1} = 0$ and $Z_{S1} = 0$, the input impedance seen by the voltage generator $V_{S2}$, with $Z_{S2} = 0$, is $Z_{in,2} = V_{S2} / I_{2,L}(0)$ or $Z_{in,2} = V_{S2} / I_{2,R}(0)$. Using (17), (18) and (21), the input impedance is $Z_{in,2} = \det(\underline{\mathbf{Z}}_{diff}) / Z_{diff,11}$, where, $Z_{diff,11}$ is the (1, 1) entry of matrix $\underline{\mathbf{Z}}_{diff}$. The resonance frequencies of the cavity can be obtained either graphically or numerically by solving $\text{Im}(Z_{in,1}) = 0$ or $\text{Im}(Z_{in,2}) = 0$ for $\omega$.

**Appendix C: Energy and power loss distributions**

The formulation to calculate the stored energy and power loss is first developed for a cavity composed of the unit cell in Fig. 6 (a) and the modification of the formulation to the cavity composed of the unit cell in Fig. 7 (a) will be discussed subsequently. We follow the notations in Sec. III and IV and the notations for MTLs in [27]-[29]. The numbering of the unit cells begin from $z = -Nd/2$ and starts from $n = 1$ ending with $n = N$. Knowing the voltage distribution $\mathbf{V}(z) = [V_1(z) \ V_2(z)]^T$ and current distribution $\mathbf{I}(z) = [I_1(z) \ I_2(z)]^T$ at any point $z$ in the cavity using formalism in Appendix B, the spatial distribution of stored time-average energy (per unit length) is given by

$$w_{em}(z) = w_e(z) + w_m(z), \tag{22}$$

where

$$w_e(z) = \frac{1}{4}\mathbf{V}(z)\underline{\mathbf{C}}(z)\mathbf{V}^*(z) \tag{23}$$

is the spatial distribution of stored time-average electric energy (per unit length) and

$$w_m(z) = \frac{1}{4}\mathbf{I}(z)\underline{\mathbf{L}}(z)\mathbf{I}^*(z) \tag{24}$$

is the spatial distribution of stored time-average magnetic energy (per unit length), respectively at any point $z$ in the cavity. Here, $\underline{\mathbf{C}}(z)$ is the capacitance matrix (per unit length) that is either $\underline{\mathbf{C}}_A$ or $\underline{\mathbf{C}}_B$, depending whether $z$ is in the segment A or B, respectively and $\underline{\mathbf{L}}(z)$ is the inductance matrix (per unit length) that is either $\underline{\mathbf{L}}_A$ or $\underline{\mathbf{L}}_B$, depending whether $z$ is in the segment A or B, respectively.

Furthermore, the time-average power lost per unit length is defined as $p_l(z) = \frac{1}{2}\mathbf{I}(z)\underline{\mathbf{R}}_s(z)\mathbf{I}^*(z)$, where $\underline{\mathbf{R}}_s(z)$ is either $\underline{\mathbf{R}}_{s,A}$ or $\underline{\mathbf{R}}_{s,B}$, depending whether $z$ is in the segment A or B, respectively. The spatial distributions of stored time-average energy (per unit length) plotted in Figs. 3 (e, f) and the spatial distributions of power loss (per unit length) in Figs. 3 (c, d) were plotted using the formulation in (22)-(24).

The total stored time-average energy in the entire cavity formed using the unit cell in Fig. 6 (a) is obtained from

$$W_{\text{stored}} = \int_{-L/2}^{+L/2} w_{em}(z)dz', \tag{24}$$

and the total power loss in the entire cavity is calculated using

$$P_{\text{lost}} = \int_{-L/2}^{+L/2} p_l(z)dz'. \tag{24}$$

The formulation describing the stored energies and power loss for the cavity composed of the unit cell in Fig. 7 (a) can be obtained by modifications to the formulation presented in the above paragraph. In particular, note that when assuming a lossless capacitive network, only the expression for the stored time-average electric energy $w_e(z)$ needs to be modified while the expressions for the stored time-average magnetic energy $w_m(z)$ and power loss $p_l(z)$ remain the same. Therefore, the formulation in (22)-(24) is still applicable but the additional stored time-average electric energy in the lumped capacitor network is accounted for by

$$W_{e,n}^{C} = \frac{1}{4} \mathbf{V}(z_{cap,n}) \begin{bmatrix} C_1 + C_m & -C_m \\ -C_m & C_2 + C_m \end{bmatrix} \mathbf{V}^{*}(z_{cap,n}), \tag{25}$$

where, $z_{cap,n}$ is the location of the lumped capacitor network in the $n^{th}$ unit cell with $1 \leq n \leq N$. In Fig. 9, the stored time-average energy (per unit length) in the TLs, which is plotted using solid black curve, was computed using formulation in (22)-(24) while the stored time-average electric energy in the lumped capacitor networks, which is plotted at discrete $z$ locations using red color, was computed using (25).

The total stored time-average energy in the entire cavity formed using the unit cell in Fig. 7 (a) is

$$W_{\text{stored}} = \int_{-L/2}^{+L/2} w_{em}(z) dz' + \sum_{n=1}^{N} W_{e,n}^{C} \tag{26}$$

while (24) is still applicable to calculate the total power loss in the entire cavity.

**Acknowledgements**

This research was supported by AFOSR MURI Grant FA9550-12-1-0489 administered through the University of New Mexico. The authors are grateful to Prof. Edl Schamiloglu from the Department of Electrical and Computer Engineering, University of New Mexico, to Dr. Michael A. Shapiro from the Plasma Science and Fusion Center of MIT, and to Dr. Guillermo Reyes from the Department of Mathematics, University of California, Irvine for helpful discussions.

**References**

[1] J. Benford, J. A. Swegle and E. Schamiloglu, "Enabling Technologies" in *High Power Microwaves*, 2$^{nd}$Edn. Boca Raton, FL: CRC Press, 2007.

[2] S. H. Gold and G. S. Nusinovich, "Review of high-power microwave source research," *Rev. Sci. Instrum.*, vol. 68, pp. 3945-3974, Aug. 1997.

[3] A. L. Vikharev, et. al, Active Compression of RF pulses in *Quasi-Optical Control of Intense Microwave Transmission*, NATO Science Series II: Mathematics, Physics and Chemistry Volume 203, J. l. Hirschfield and M. I. Petelin, Eds. Dordrechet, The Netherlands: Springer, 2005.

[4] R. A. Alvarez, "Some properties of microwave resonant cavities relevant to pulse-compression power amplification,"*Rev. Sci. Instrum.*, vol. 57, pp. 2481-2488, Jun. 1986.

[5] R. A. Alvarez, D. P. Byrne, and R. M. Johnson, "Prepulse suppression in microwave pulse-compression cavities," *Rev. Sci. Instrum.*, vol. 57, pp. 2475-2480, Jun. 1986.

[6] V. A. Avgustinovich, S. N. Artemenko, V. L. Kaminsky, S. N. Novikov, and Yu. G. Yushkov, "Note: Resonant microwave compressor with two output ports for synchronous energy extraction," *Rev. Sci. Instrum.*, vol. 82, 046104, Apr. 2011.

[7] A. Shlapakovski, S. Artemenko, P. Chumerin and Yu.Yushkov, "Controlling output pulse and prepulse in a resonant microwave pulse compressor," *J. Appl. Phys.*, vol. 113, 054503, Feb. 2013.

[8] A. D. Andreev et al., "A simplified theory of microwave pulse compression," *Circuit and Electromagnetic System Design Notes*, note 57, Aug. 2008, www.ece.unm.edu/summa/notes.

[9] E. G. Farr et al., "Microwave pulse compression experiments at low and high power," *Circuit and Electromagnetic System Design Notes*, note 63, Jan.2010, www.ece.unm.edu/summa/notes.

[10] S. G. Tantawi, R. D. Ruth and A. E. Vlieks, "Active radio frequency pulse compression using switched resonant delay lines," *Nuclear Instruments and Methods in Physics Research A*, Vol. 370, pp. 297-302, Feb. 1996.

[11] J. Guo and S. Tantawi, "Active RF pulse compression using an electrically controlled semiconductor switch," *New J. Phys.,* Vol. 8, pp. 293-310, Jun. 2006.

[12] O. A. Ivanov, et. al, "Active quasioptical Ka-band RF pulse compressor switched by a diffraction grating," *Phys. Rev. ST Accel. Beams*, Vol. 12, 093501, Sept. 2009.

[13] P. Paulus, L. Stoll and D. Jager, "*Optoelectronic pulse compression of microwave signals*," IEEE Trans. Microwave Theory Technol., Vol. 35, No. 11, pp. 1014-1018, Nov. 1987.

[14] V. A. Avgustinovich, S. N. Artemenko, and A. S. Shlapakovski, "Resonant frequency-tunable microwave compressors," *J. Commun. Technol. Electron.*, Vol. 54, No. 6, pp. 721-732, Jun. 2009.

[15] A. Vikharev, et. al, "High power active X-band pulse compressor using plasma switches," *Phys. Rev. ST Accel. Beams*, Vol. 12, 062003, Jun. 2009.

[16] D. M. Pozar, *Microwave Engineering*, 2$^{nd}$Edn. New York, NY: John Wiley and Sons, 1998.

[17] R. E .Collins, *Foundations for microwave engineering*, 2$^{nd}$Edn. New York, NY: John Wiley and Sons, 2001.

[18] A. Figotin and I. Vitebsky, "Gigantic transmission band edge resonance in periodic stacks of anisotropic layers,"*Phys. Rev. E*, vol. 72, 036619, Sept. 2005.

[19] A. Figotin and I. Vitebsky, "Frozen light in photonic crystals with degenerate band edge," *Phys. Rev. E*, vol. 74, 066613, Dec. 2006.

[20] A. Figotin and I. Vitebsky, "Slow wave phenomena in photonic crystals," *Laser Photon Rev.*, vol. 5, no.2, pp. 201-213, Mar. 2006.

[21] K. Y. Jung and F. L. Teixeira, "Photonic crystals with a degenerate band edge: Field enhancement effects and sensitivity analysis," *Phys. Rev. B*, vol. 77, 125108, Mar. 2008.

[22] C. Locker, K. Sertel and J. L. Volakis, "Emulation of propagation in layered anisotropic media with equivalent coupled microstrip lines," *IEEE Microw. Wireless Compon. Lett.*, vol. 16,  no. 12,  pp.642 -644, Dec. 2006.

[23] G. Mumcu, K. Sertel, and J. L. Volakis, "Lumped Circuit Models for Degenerate Band Edge and Magnetic Photonic Crystals," *IEEE Microw. Wireless  Compon. Lett.*, vol.20, no.1, pp.4-6, Jan. 2010.

[24] N. Marcuvitz, *Waveguide Handbook*, Radiation Laboratory Series, Vol. 10, McGraw-Hill Book Company, Inc., New York, 1950.

[25] N. Marcuvitz and J. Schwinger, "On the Representation of the Electric and Magnetic Fields Produced by Currents and Discontinuities in Wave Guides. I,"*J. Appl. Phys.*,vol. 22, pp. 806-819, Jun. 1951.

[26] L. B. Felsen and N. Marcuvitz, *Radiation and scattering of waves*, New York, NY: IEEE Press, 1994.

[27] C. R. Paul, *Analysis of Multiconductor Transmission Lines*. 2nd Ed. Hoboken, NJ: John Wiley & Sons, 2008.

[28] C. R. Paul, "Decoupling the multiconductor transmission line equations," *IEEE Trans. Microw. Theory Tech.*, vol. 44, no. 8, pp.1429 -1440, Aug. 1996.

[29] C. R. Paul, "A brief history of work in transmission lines for EMC applications," *IEEE Trans. Electromagn. Compat.*, vol. 49,  no. 2,  pp.237 -252, May 2007.

[30] L. B. Felsen and W. K. Kahn, "Transfer characteristics of 2N-port networks" in *Proceedings of the symposium on Millimeter Waves, New York-1959*, J. Fox (Eds).  Brooklyn, NY: Polytechnic Press, 1960.